\documentstyle[preprint,aps]{revtex}
\begin{document}
\begin{center}
\begin{Large}
{\bf The band structure of MgB$_{2}$ with different lattice constants}
\end{Large}

\begin{large}
Xiangang Wan, Jinming Dong, Hongming Weng and D. Y. Xing 
\end{large}
  
Group of Computational Condensed Matter Physics, National
Laboratory of Solid State Microstructures and Department of Physics, Nanjing
University, Nanjing 210093, P.R.China
\end{center}

\begin{abstract}
We report a detailed study of the electronic structure of the MgB$_{2}$ with
different lattice constants by using the full-potential linearized augmented 
plane wave(FPLAPW) method. It is found that the lattice parameters have great
effect on the $\sigma$ band of Boron. Our results indicate
that increasing the lattice constant along the
{\it c} axis will increase the density of states(DOS) at the Fermi
level, shift the $\sigma$ band upward, and increase the hole
number in the $\sigma$ band. 
So, the superconducting transition temperature T$_{c}$ will be raised
correspondingly. 
Changing the lattice constant along {\it a} axis has the opposite effect to
that of the {\it c} axis. Our result is in agreement with experiment. 
A possible way of searching for higher T$_{c}$ superconductor has been
indicated, i.e., making MgB$_{2}$ to have longer {\it c} axis and shorter 
{\it a, b} axis by doping.
\end{abstract}

\vskip 2cm


\newpage  
  
The discovery of superconductivity in magnesium diboride (MgB$_{2}$) [1,2]
has stimulated world wide excitement. The superconducting
transition temperature for MgB$_{2}$ ($T_C \approx 39$ K)
exceeds by almost two times the record values of $T_C$ for conventional
$B1$- and $A15$-type intermetallic superconductors (SC) [3],
which is by far the highest if we exclude copper
oxides and $C_{60}$ based materials. As distinct
from the high-temperature superconductor, MgB$_{2}$ has an exclusively
simple composition and crystal structure [4].
The B isotope shift of T$_c$ reported by Bud'ko [5]
and most other early experimental data[6]
suggests conventional BCS strong-coupling s-wave electron-phonon (EP) pairing.

Band structure calculations show
that the compound is not only quite
ionic with a reasonable DOS, but also has strong
covalent B-B bonding (the bonding-antibonding splitting due to
in-plane B-B hopping is about 6 eV)
and thus exhibits strong electron-phonon interactions[7, 8, 9].
Mg $s$-states are pushed up by the B $p_{z}$ orbitals and
fully donate their electrons to the boron-derived conduction bands.
Hirsch used a hole-superconductivity model to explain the high temperature
superconductivity in MgB$_{2}$ as driven by undressing of hole carriers in
the planar boron p$_{x,y}$ orbitals in the negatively charged B$^{-}$
planes [10]. The model by Hirsch predicts a positive pressure effect on
T$_{c}$. However, high pressure studies show negative pressure coefficient
of T$_{c}$[11, 12]. Based on an estimate of phonon
frequencies and band structure calculations, Kortus et al[8]
explain the superconductivity in $MgB_2$ as a result of strong
electron-phonon coupling and An and Pickett[7] attributed it to
the behavior of $p_{x,y}$-band holes in negatively charged boron planes.
Many authors[7-9] emphasize the significant role
of metallic B states in the appearance of superconductivity.
According to the McMillan formula for $T_C$[13], the high transition temperature 
is probably due to a high density of states at the Fermi
level, $N(E_F)$, high averaged electron-ion matrix elements,
as well as high phonon frequencies, which increase
for light elements and depend on $B-B$ and $M-B$ bonding.

Past experience said that T$_{c}$ should be strongly affected by doping
or changing the lattice constants.
Some experiments concerning substitutions on the Mg
site have been done[14, 15]. Substitution of Al leads, apart from doping with
electrons, to a compression of the structure due to the difference in
ionic radius between Al and Mg. The compression is anisotropic and an Al
content of x=0.1 leads to a structural instability. The rate of compression
along the {\it c} axis is about twice as much as along the {\it a} axis[14].
Both hole doping by substitution of Li$^{+}$[15], 
and electron doping by substitution of Al$^{3+}$[14] for
Mg$^{2+}$ led to a decrease of T$_{c}$. In fact, Al doping will destroy bulk
superconductivity when the Al content x is larger than 0.3[14].
There are also many experiments concerning Boron sublattice doping[16]. The
lattice parameter {\it a} decreases almost linearly with increasing carbon content
x, while the {\it c} parameter remains unchanged, indicating that carbon is
exclusively substituted in the Boron honeycomb layer without affecting the
interlayer interactions. T$_{c}$ also decreases
about linearly as a function of the carbon concentration.
A recent experiment had also found
BiB$_{2}$ not to be a superconductor[17]. BiB$_{2}$ and MgB$_{2}$ are
isostructural, and their valence electron number is the same.
So, the basic change is due to structural factors,
i.e. lattice parameter, {\it a}, and interatomic distance, {\it c/a}.
Recently, high pressure studies show that T$_{c}$ will decrease with increasing
pressure[11, 12]. Compression will decrease both lattice constant {\it a}
and {\it c}, but it is not yet know
which one of them is responsible for decreasing T$_{c}$.
So, it is very interesting to investigate the
Boron band structure, especially the $\pi$ and $\sigma$ band, 
as a function of lattice constants {\it a} and {\it c}.

We have used the highly accurate all-electron full-potential linear
augmented plane wave method[18, 19]. The standard local density
approximation to the electron exchange-correlation potential was used
together with the generalized gradient corrections of Perdew[20].
The muffin-tin sphere radii (R) of 2.00 a.u. and 1.50 a.u. were chosen for
the Mg and B atoms, respectively, with a cutoff RK$_{max}$=8.0.

MgB$_{2}$ is isomorphous with AlB$_{2}$[4]; the lattice constants of the
hexagonal unit cell are a$_{0}$=3.0834 and c$_{0}$=3.5213 $\AA$. 
From the existing experiments we can see that high hydrostatic pressure has
an anisotropic influence along {\it a} and {\it c}. The compression along
the {\it c} axis is 64$\%$ stronger than along the {\it a} axis, which is in line
with the weaker Mg-B bond[21]. Thermal compression along the {\it c} axis is
about twice the one along the {\it a} axis. So,
we fix {\it a}=3.0834$\AA$, and vary lattice constant {\it c}. The total 
energy-vs-c relation is shown in Fig.1. It can be seen that {\it c}={\it
$c_{0}$}=3.5213 
$\AA$ with the lowest total energy agrees with the experimental result.

The band structure of MgB$_{2}$ is shown in Fig.2. 
For convenience, we set E$_{f}$=0. From it we can see that
there are two kinds of bands: the $\sigma$ and $\pi$ band. Both of them are
contributed by Boron. The $\sigma$ band along $\Gamma$-A is double
degenerate.
Change of lattice constant will not change the symmetry, and so will not change
its character. We found when {\it c}=1.4{\it c}$_{0}$, 
the $\sigma$ band has a
very small dispersion along $\Gamma$-A. But it will increase 
with decreasing {\it c}. Thus
increasing {\it c} will strengthen the 2D character of the $\sigma$ band,
and if {\it c} approaches
$\infty$, MgB$_{2}$ becomes an ideal 2D B$_{2}$ layer.
Another evident change is that
the $\sigma$ band will shift upward with respect to the 
E$_{f}$ when we increase {\it c}.
For c=0.8c$_{0}$, the $\sigma$ band is below the
Fermi energy, and the $\sigma$ bonding state would be completely filled.
For {\it c}=0.9c$_{0}$, the
$\sigma$ band will cross the Fermi energy at the $\Gamma$ point. With
increasing {\it c}, the $\sigma$ band will shift upward, and so
will have more holes. On the other hand, increasing
{\it c} will decrease the hopping coefficient t$_{pp\sigma}$ in Kortus's TB
model[8] for the $\pi$ band, 
and will cause the $\pi$ band dispersion along $\Gamma$-A
decrease. Comparing with E$_{f}$, increasing {\it c} will decrease the
position of bonding $\pi$ band at M and $\Gamma$ point, and will raise the
position of bonding $\pi$ band at A and L point.

We have also studied the effect of lattice constant {\it a} on 
the electronic structure, and
found it to be opposite to {\it c}. 
With decreasing {\it a}, 
both bonding and antibonding $\sigma$ band will move   
upward with respect to E$_{f}$, 
and the splitting between bonding and antibonding $\sigma$ band will
increase. When {\it c}=0.8{\it c$_{0}$}
and {\it a}={\it a}$_{0}$, the bonding $\sigma$ band is below the Fermi energy. Fixing 
{\it c} and decreasing the {\it a}, the bonding $\sigma$ band will move upward. From
Fig.2(d), we can see that, when {\it c}=0.8{\it c}$_{0}$ and {\it a}=
0.9{\it a}$_{0}$, its $\sigma$ band lies higher than it in Fig.2(a), and
there are holes in the bonding $\sigma$ band. The increase of splitting between 
bonding and antibonding $\sigma$ band is due to
increase of the hopping coefficient with decreasing {\it a}.
The possible reason why $\sigma$ band moves upward is electrostatic
effect. Being different from $\sigma$ band,
the bonding $\pi$ band will move downward with respect to the E$_{f}$ when
{\it a} decreases.
Compression will decrease both the {\it a} and {\it c}. But $\sigma$ bond of B-B
is much stronger than the Mg-B bond, so, it is natural to expect that
compression will mainly cause the decrease of {\it c}[21].
Thus, we can conclude that
compression will shift the $\sigma$ band downward and decrease the
number of holes in the $\sigma$ band.

We have also studied the relationship between the density of states
at the Fermi level(N$_{f}$) and the lattice constants, which is shown in Fig.3.
It is found that N$_{f}$ will increase with increasing of the 
lattice constant {\it c}, which is
in agreement with other theoretical results[22, 23]. 
Many theories and experiments show that the MgB$_{2}$ is a BCS-like
superconductor. So the
electron-phonon coupling constant $\lambda$, which enters the BCS equation,
is very important[13], and is proportional to N$_{f}$. 
Since N$_{f}$ will decrease for smaller {\it c},
T$_{c}$ will be decreased, which is in agreement with experiments[11, 12]. 
The states at the Fermi level, responsible for superconductivity,
show two different orbital characters: $p-\sigma$ bonding
(column--like FS around $\Gamma-A$)
and $p_z$, which has $\pi$-bonding and antibonding characters on the basal and
on the top ($k_z=\pi/c$) planes, respectively.  Which one of them is the
most important for the superconductivity is not yet known. In order to answer
this question, we calculate the partial DOS of $\pi$ and $\sigma$ band of MgB$_{2}$
with different lattice constants {\it c}. 
We find that with a larger {\it c},
N$_{f}$($\sigma$) will increase, whereas N$_{f}$($\pi$) will
decrease. Here, the N$_{f}(\sigma)$ (N$_{f}(\pi)$) means the density of
$\sigma$ ($\pi$) state at Fermi energy.
So, we can say that the $\sigma$ band of Boron plays a more
important role for superconductivity than the $\pi$ band. 
The p$_{\sigma}$ and p$_{\pi}$ bands move with respect to each other
with increasing {\it c}, thus inducing charge transfer between these two
bands. It is the reason why the N$_{f}(\sigma)$ has an opposite trend to
N$_{f}(\pi)$. Our result is in agreement with Goncharov and Bud'ko et al.,'s
result[24].

In summary, we calculated the electronic structure of MgB$_{2}$ for
different lattice constants. Our results show that
increasing lattice constant {\it c} will increase the DOS at the Fermi
level, shift the $\sigma$ band upward, and thus increase the hole number
in the $\sigma$ band. 
Decreasing {\it c} will lower the T$_{c}$. 
The effect of changing {\it a} is opposite to {\it c}. 
For constant {\it c}, a shorter {\it a} will raise T$_{c}$.
The main influence of compression is to decrease {\it c}, and thus
will decrease the T$_{c}$, which is in agreement with the experiments[11,
12]. A search for a higher T$_{c}$ superconductor 
by doping MgB$_{2}$, would require a longer lattice constant {\it c} and 
shorter {\it a, b}.
A possible way would be a suitable substrate, which forces MgB$_{2}$
film to have longer {\it c} but shorter 
{\it a, b}.

\begin{large}
\begin{center}
{\bf V. Acknowledgments}
\end{center}
\end{large}
  
The authors acknowledge support in this work by a grant for State Key
Program of China under Grant No. G 1998061407. The numerical
calculations in this work have been done on the SGI Origin - 2000 computer in the Group of 
Computational Condensed Matter Physics, 
in the National Laboratory of Solid State Microstructures, Nanjing University.

\newpage
\begin{center}\begin{large} {\bf Figure Captions }
\end{large}\end{center}

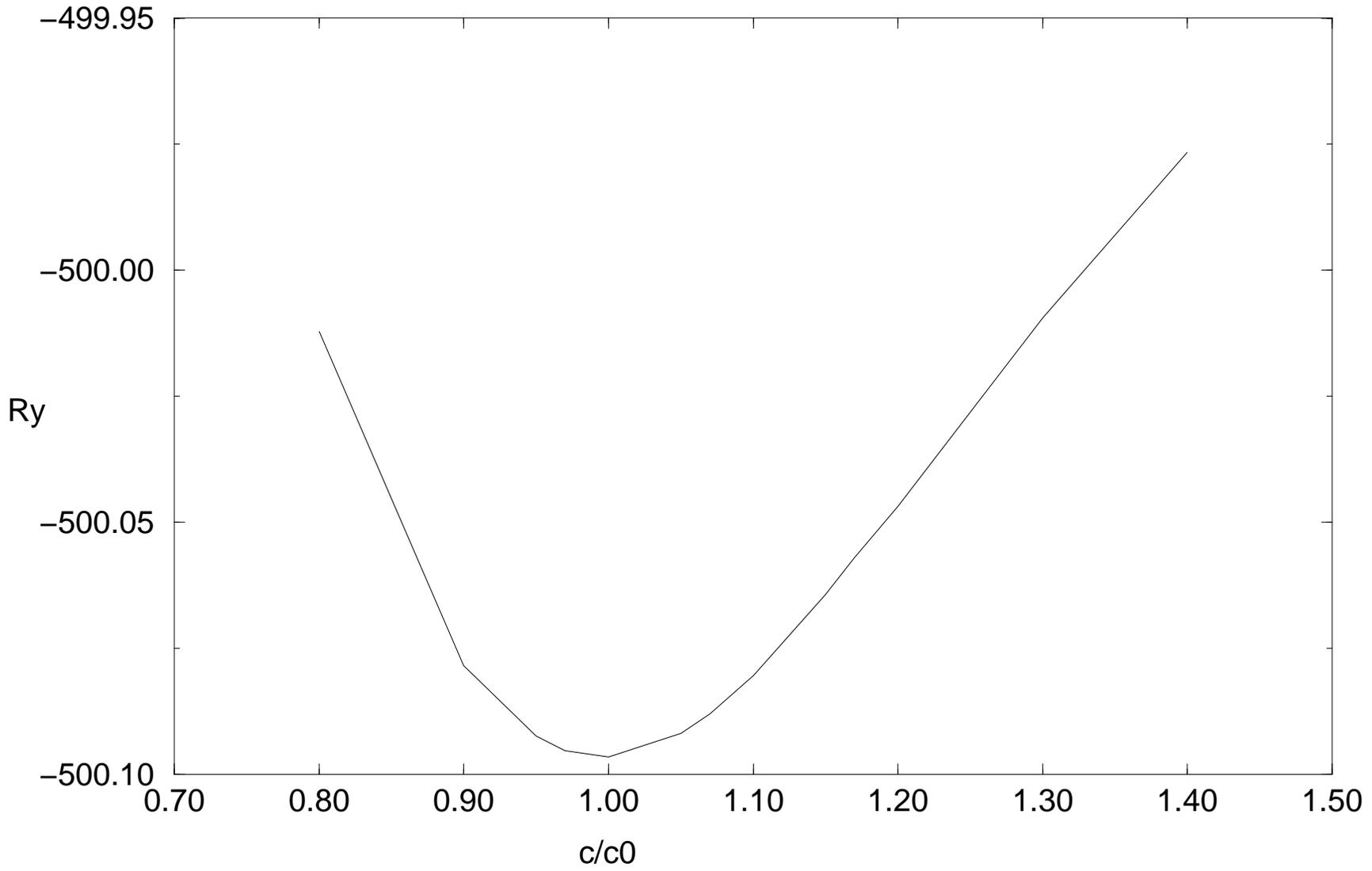
{\bf Figure 1} The total energy of MgB$_{2}$ as a function of lattice
parameter {\it c/c$_{0}$}. {\it $c_{0}$}= $3.5213\AA$.

{\bf Figure 2} The band structure of MgB$_{2}$ with different lattice
constants. The circles represent the Boron $\sigma$ band.
{\it $a_{0}$}= $3.0834\AA$,
{\it $c_{0}$}= $3.5213\AA$.
(a) {\it a}={\it a}$_{0}$, {\it c}=0.8{\it c}$_{0}$;
(b) {\it a}={\it a}$_{0}$, {\it c}={\it c}$_{0}$;
(c) {\it a}={\it a}$_{0}$, {\it c}=1.4{\it c}$_{0}$;
(d) {\it a}=0.9{\it a}$_{0}$, {\it c}=0.8{\it c}$_{0}$.

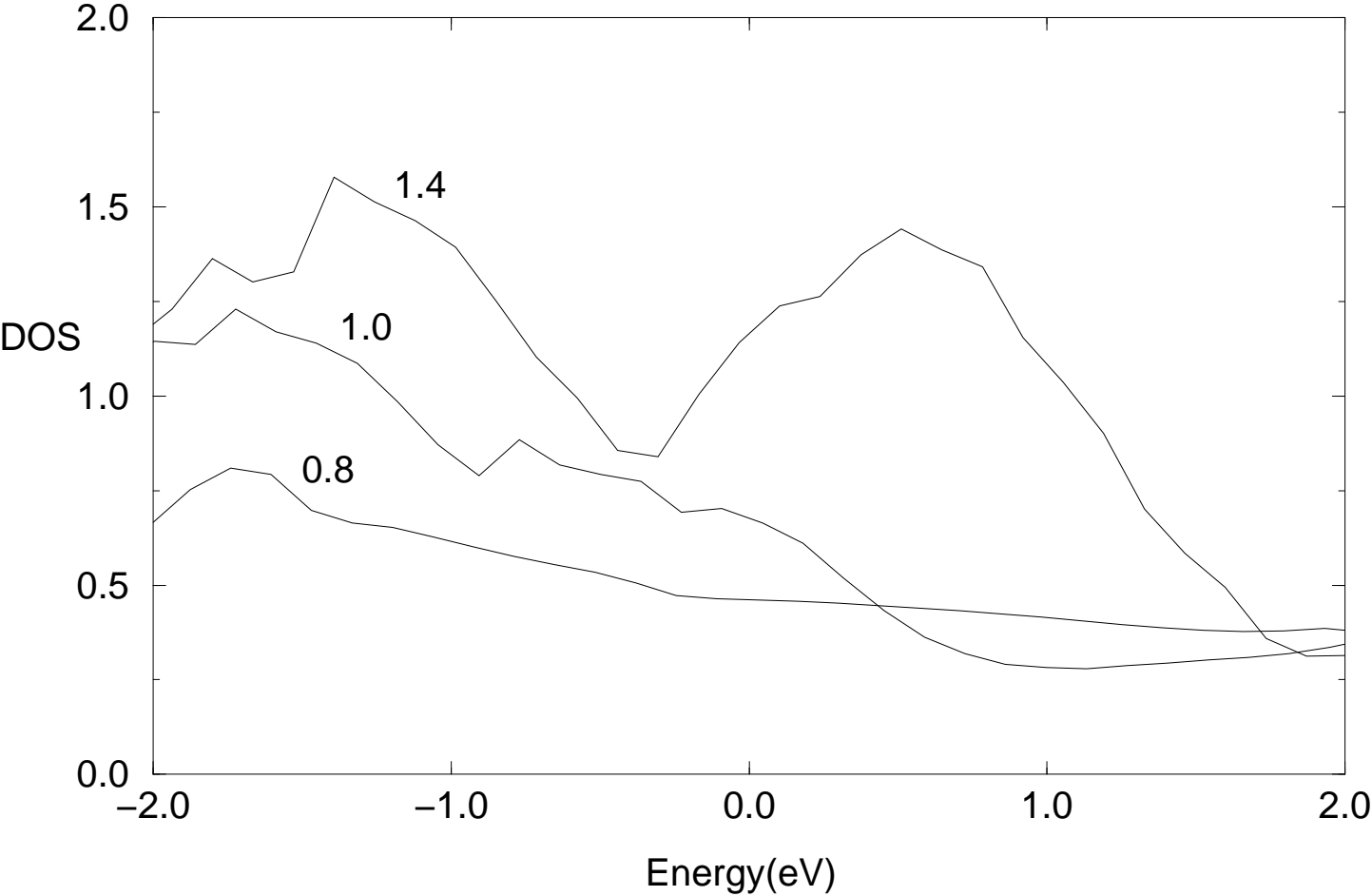
{\bf Figure 3} Density of states near E$_{f}$ for MgB$_{2}$
with different lattice constants, {\it c}/{\it c}$_{0}$=0.8, 1.0 and 1.4.
Fermi level is set at the zero.
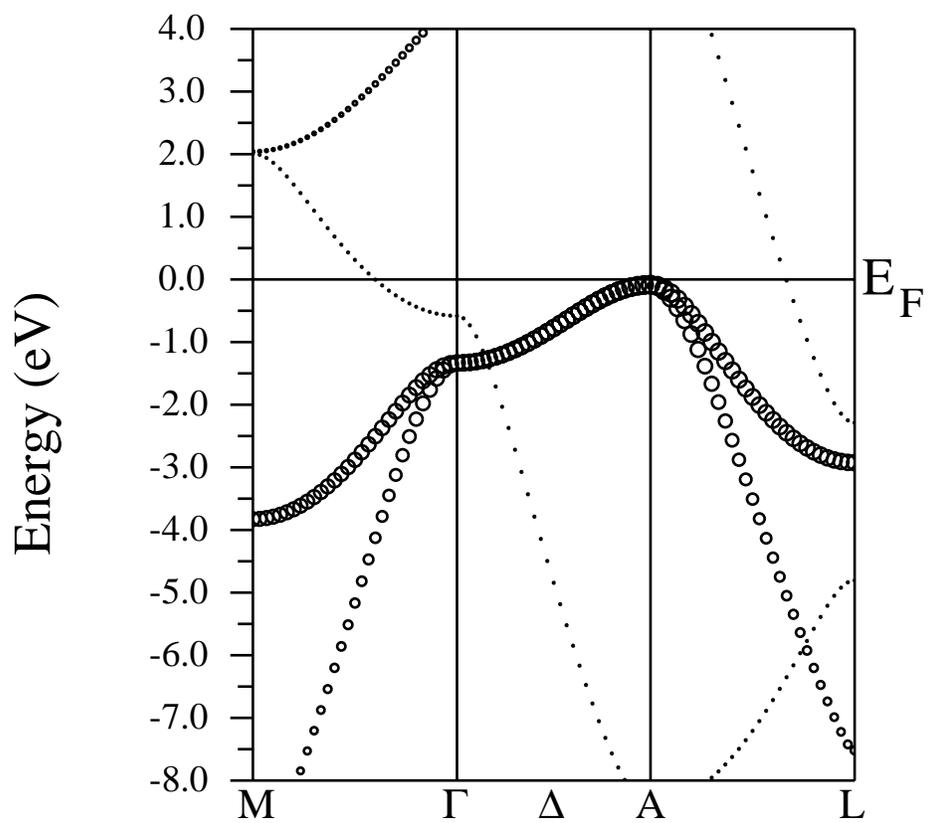
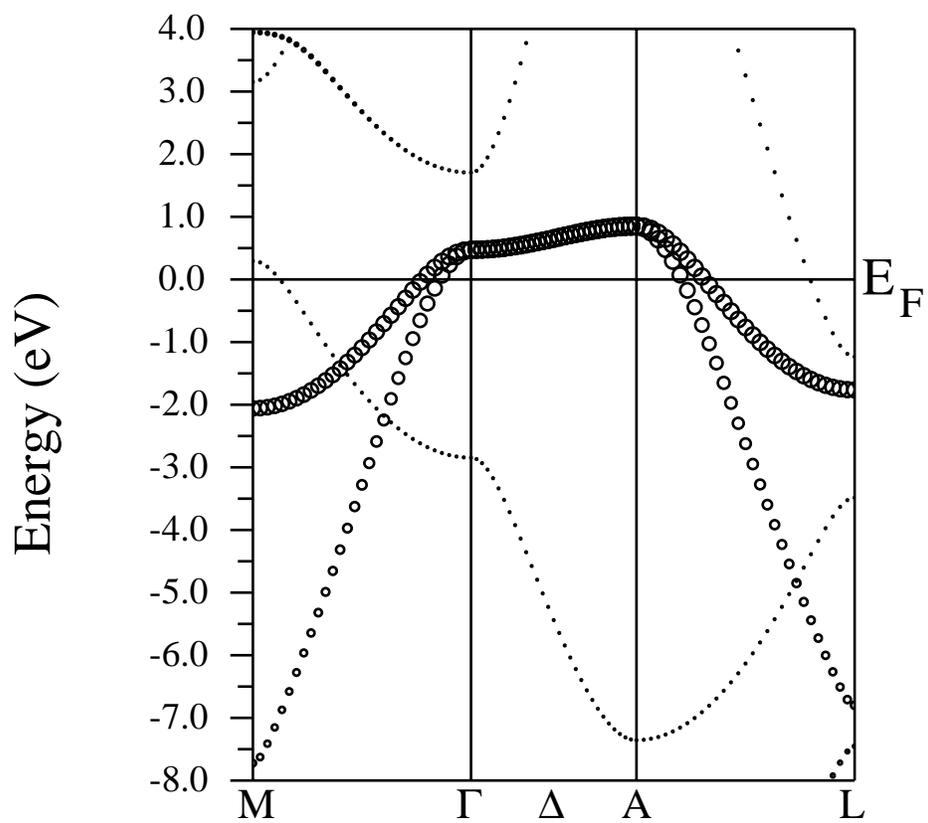
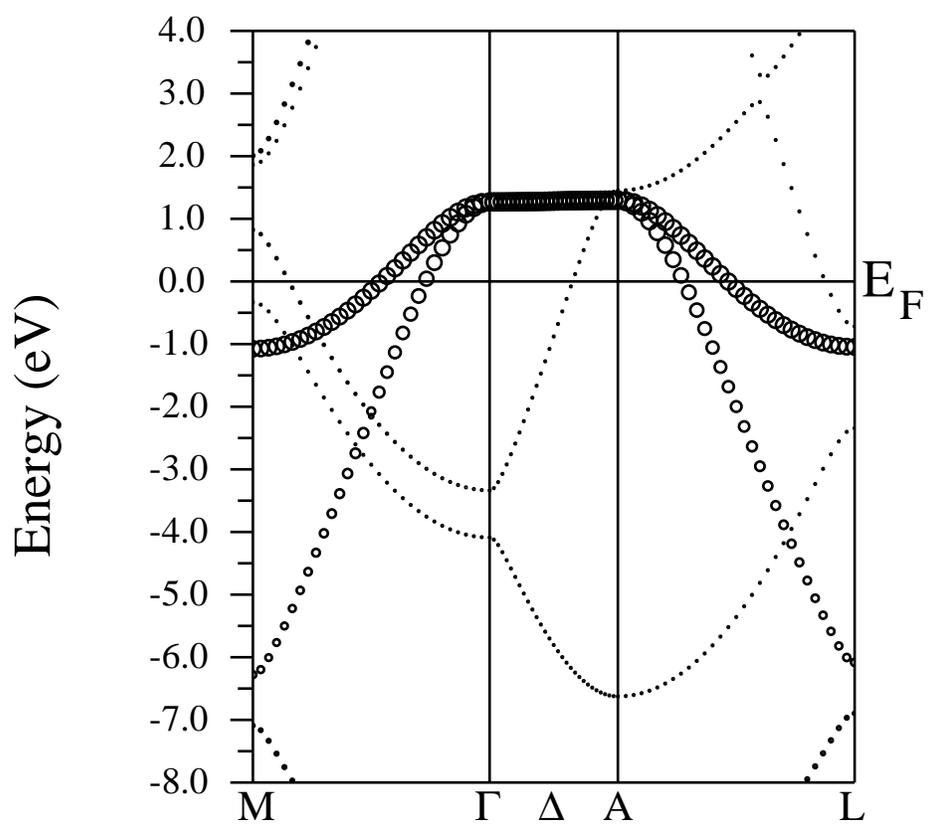
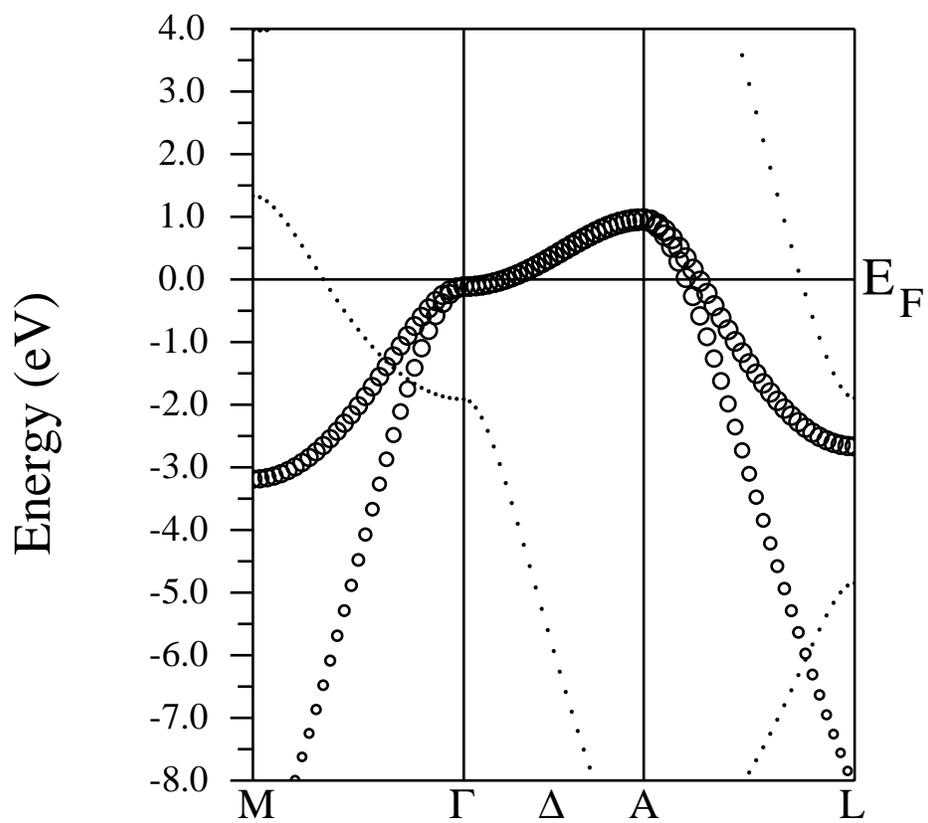
\end{document}